%% file: main.tex
\documentclass[conference]{IEEEtran}
\IEEEoverridecommandlockouts
\usepackage{cite}
\usepackage{amsmath,amssymb,amsfonts}
\usepackage{algorithmic}
\usepackage{graphicx}
\usepackage{textcomp}
\usepackage{xcolor}
\usepackage{multirow}
\usepackage{comment}
\usepackage [ short ]{ optidef }
\usepackage[hyphens]{url}
\usepackage[hidelinks]{hyperref}
\hypersetup{breaklinks=true}
\usepackage[justification=centering]{caption}
\usepackage{caption}
\usepackage{adjustbox}
\usepackage{subcaption}
\usepackage{stfloats}
\usepackage{makecell}
\usepackage{rotating,lipsum}
\def\BibTeX{{\rm B\kern-.05em{\sc i\kern-.025em b}\kern-.08em
    T\kern-.1667em\lower.7ex\hbox{E}\kern-.125emX}}
\begin{document}

\title{Electric Vehicle Integration using Large-Scale Combined Transmission and Distribution Grid Models\\
}


\author{\IEEEauthorblockN{Diana Wallison, Lyric Haylow, Jessica Wert, \\ Jonathan M. Snodgrass, Thomas J. Overbye\\  }
\IEEEauthorblockA{\textit{Department of Electrical and Computer Engineering} \\
\textit{Texas A\&M University}\\
College Station, TX \\
\{diwalli, lyrich237339, jwert, snodgrass, overbye\}@tamu.edu}
\and
\IEEEauthorblockN{Yanzhi (Ann) Xu}
\IEEEauthorblockA{\textit{ElectroTempo} \\
Arlington, VA \\
 \{ann.xu\}@electrotempo.com}

}
\maketitle

\begin{abstract}
In this paper, we propose a unifying co-simulation framework integrating transportation demand, grid assets, land use, demographics, and emissions to optimally accelerate electric vehicle (EV) development as well as measure the impact of EV integration. 96 urban and long-haul truck charging demand simulations were developed and integrated into a combined transmission and distribution (T\&D) simulation, encompassing the Houston/Dallas/Fort Worth area. The T\&D scenarios are then used to develop cost optimization strategies to determine optimal placement and sizing of truck charging infrastructure that minimize infrastructure costs.

\end{abstract}

\begin{IEEEkeywords}
Electrification, Electric Vehicles, Trucks, Transportation, Transmission, Distribution, Charging, ac Optimal Power Flow
\end{IEEEkeywords}

\section{Introduction} \label{section:Introduction}
\input{Sections/I_Introduction}

\section{Modeling EV Charging Demand}
\input{Sections/II_Modeling} \label{section:Modeling}

\section{Mapping EV Charging Demand}
\input{Sections/Mapping} \label{section:Mapping}

\section{Co-Simulation} \label{section:CoSimulation}
\input{Sections/III_CoSimulation}

\section{Case Study} \label{section:CaseStudy}
\input{Sections/IV_Case_Study}

\section{Results and Discussion} \label{section:Results}
\input{Sections/V_Results_Discussion}

\section{Conclusions}
\input{Sections/VII_Conclusion}

\section*{Acknowledgment}
This work is supported through funding provided by the U.S. Department of Energy (DOE) Office of Energy Efficiency and Renewable Energy (EERE) Grant No. DE-EE0009665, administered through a sub-award from ElectroTempo. The authors gratefully acknowledge ElectroTempo for the EV charging datasets, and their guidance on the project.

\Urlmuskip=0mu plus 1mu\relax
\bibliographystyle{IEEEtran}
\bibliography{bibi.bib}

\end{document}

%% file: Sections/I_Introduction.tex
Utilities and grid operators such as RTOs and ISOs generally plan and operate their distribution and transmission systems independently, with separate operating centers and planning approaches.
In these cases distribution load is often assumed to be "ideal" meaning a constant power balanced 3-phase load; while the transmission system is generally modeled using an "infinite bus" at rated nominal voltage. 

EV integration studies are also often overly simplified, using models that neglect either the transportation or electric grid dynamics alongside detailed modeling. Studies usually oversimplify their transportation grid models. There are also few large scale combined T\&D system studies as most researchers use small transmission or distribution models. 

In Reference \cite{cosim2}, the spatial aspect of EV integration is accounted for by associate traveling purpose with nodal load types in a distribution system, however the transmission system is not modeled leaving the impact of EV charging on the full power system unexplored. In contrast, reference \cite{cosim3} models both the transmission and distribution system. The model, however, only includes one distribution feeder and a IEEE 9-bus transmission system, meaning that the co-simulation's potential for scaling for larger systems is not necessarily assured. This issue persists in references \cite{cosim6} and \cite{cosim7}. Reference \cite{cosim6} only models an IEEE 9 bus system with only three load buses and only one distribution feeder. The co-simulation in reference \cite{cosim7} is even smaller, with only 20 households included in its electrical power grid.

There are similar issues with the studies in references \cite{cosim4} and \cite{cosim5}. Both references only model the distribution system and have small case studies, with the distribution system used in reference \cite{cosim4} being a IEEE 37 Node Test Feeder, and the distribution system used in \cite{cosim5} being a IEEE 33 node test system.

Studies that have large detailed distribution systems and study the impacts of EV charging, however usually model the transmission and distribution separately or only model the distribution system. Reference \cite{cosim_new} uses a large scale distribution system in California to study grid congestion, however it does not present data on the effect of congestion on the transmission system.

In general, from previous EV integration optimization studies the transmission system is often ignored entirely, not allowing for the full impact of EV charging demand to be studied from a complete bottom-up perspective. In contrast to these previous studies, our study covers 2,304 different EV charging snapshots, compiled from 96 scenarios, in a 24 hour period. The novel contributions of the study include: 
\begin{itemize}
    \item The extensive power system and transportation data, acquired from the 96 charging scenarios
    \item Conducting these extensive studies on a large scale system with over 3,000,000 distribution nodes
    \item An optimization strategy that determines optimal control of the transmission Grid, with co-simulation showing the effects on both the transmission system and distribution systems down to the distribution node level
\end{itemize}

%% file: Sections/II_Modeling.tex
\subsection{Modeling EV Charging Demand}
\label{subsec:EVCharging}

\subsubsection{Traffic Modeling}
The traffic modeling for this paper is performed using a regional Travel Demand Model (TDM). Regional transportation planning processes often use TDMs in their calculations, to estimate energy consumption for EVs. For this paper, energy consumption is estimated using trip information generated by the TDM including: origins and destinations with their exact geographical coordinates, vehicle type, trip distances, duration, and speeds for various time periods in a day.

\subsubsection{EV Charging Load Modeling}

The spatio-temporal charging demand developed utilizing the transportation model is the main connection point between the transportation model and the T\&D power system.

The travel model, discussed previously, is combined with a vehicle dynamics model and surveys on travel and charging behaviours, to calculate EV energy consumption. 
Survey data is used to determine trip duration and locations. This survey data contains personal interviews that measure travel patterns of vehicles entering and exiting certain areas \cite{TravelSurvey}. combining this data with trip mileage, vehicle registration data, and vehicle emissions data allows us to calculate the overall charging demand \cite{annref4}.

This process goes as follows, first, the vehicle type is determined, then the model vehicle attributes, including vehicle class and type, are assigned and used to estimate the energy consumption of each trip. The attributed of the trip are considered so that demand can be assigned to specific locations at specific times withing the transportation network.

Each trip has an energy usage associated with it leading to an estimated hourly charging demand within the transportation network. This estimated demand is calculated based on the vehicle atributes described previously and the modeled miles traveled by the vehicle. Uncertainties within the model are modeled using Markov chain Monte Carlo (MCMC) \cite{wang2018markov}.

Public charging, workplace charging, and home charging are all modeled separately using different charging logics. Public charging is defined, in our studies, as energy demanded from trips that do not originate from home  and have a non-work related purpose. If the energy requirement for a trip will bring the vehicle below a cut-off point of charge, it is assumed that the vehicle will charge between destinations in a multi-destination trip, even if the charge will not necessarily be full-recharge for the vehicle. On trips such as these, it is assumed the vehicle will charge using fast chargers at 100 kw. 
Each trip has a threshold for charging range anxiety that is treated as a random variable with a gamma distribution, estimated using a stated preference survey, to account for differences in traveler behavior. The mean range anxiety estimated was 32.9 miles.The public charging demand is estimated using a function of the vehicle's consumption rate and the mileage of the previous trip.
When a trip is set to end at a workplace, all charging demand is assigned to the end of the trip as workplace charging, meaning the vehicle will charge to full at 19kw.
 
 Finally, we assume the vehicle charges overnight to full at 2.4 kW immediately after arrival at home (the end time of the last trip of the day). The random generation of EVs by household, assignment of EV types to trips, and the resulting energy demand assignment within the network are repeated to produce an average estimate of charging demand. Our other assumptions are that the first trip of the day probability is not correlated with location; distance traveled is not correlated with location or EV type; and that range anxiety is not correlated with time, location, or EV type.

It is assumed, for our study, that short-haul trucks, with a range of 350 miles, charge at their depots, using 100 kw level 3 (fast) chargers. The fast-charging demand forecasting model based on a data-driven approach and human decision-making behavior is described in a previous work \cite{xing2020urban}. The required locations for these chargers are determined by commercial vehicle data \cite{xu2020scalable}. 

The ‘Electric Vehicles–Power Grid–Traffic Network’ fusion architecture is constructed using data  mining alongside established models. One key aspect of this architecture is the depot probability model. This model is applied to every land parcel, in the region, and predicts it's possibility of being a depot. The model was trained on 1,000 land parcels using satellite imagery. This step is crucial to identify locations within the established transportation network where a large truck would be able to charge, as they require specialized charging stations different from the standard light duty EV chargers.

 This model reflects the complex correlation structure between numerous attributes of vehicle trips, including vehicle class and type, fuel type, cargo type, origin county, trip time of day, and whether the trip is the first or last of the day. Whether a trip is the last of the day is important as it determines the assumed charging behavior for the EV. If the current trip is the final trip of the day, the model assumes trucks will be charged overnight, to as close to full as can be achieved before the scheduled first trip of the next day. For trips that are not the last trip of the day, the mileage of the trip and EV truck energy consumption rates used to calculate the charging demand. Any unmet demand from the day, if the trip does not end within 100 meters of a depot is distributed proportionately to all nodes that contain depots for overnight charging.

\subsection{Mapping EV Load to the Distribution System}
With information on geographical coordinates of the charging demand and geographic topology of the electric grid, the EV charging loads are mapped to the appropriate distribution-level nodes using the mapping methodology developed in our previous work \cite{ISGT2021_coupled}. The mapping takes the latitudes and longitudes of nodes in the distribution system and creates tessellating service areas using a Voronoi diagram. The nodes are centrally-located within their respective service territories. As this work includes distribution grid simulation, the service areas are needed at the distribution system level, which is connected to the transmission grid. 

The process of this mapping for purely distribution-level simulation is as follows:

\begin{enumerate}
    \item Using the geographic feeder node coordinates, create a Voronoi diagram to represent the division of service areas for each distribution feeder node.
    \item For each EV charging station, determine the feeder node corresponding to the service area in which the charging station lies. 
    \item Include a load in the electric grid model to represent the aggregate EV charging stations within the service area. 
\end{enumerate}

\begin{figure*}[htbp]
    \centering
    \includegraphics[width=2\columnwidth]{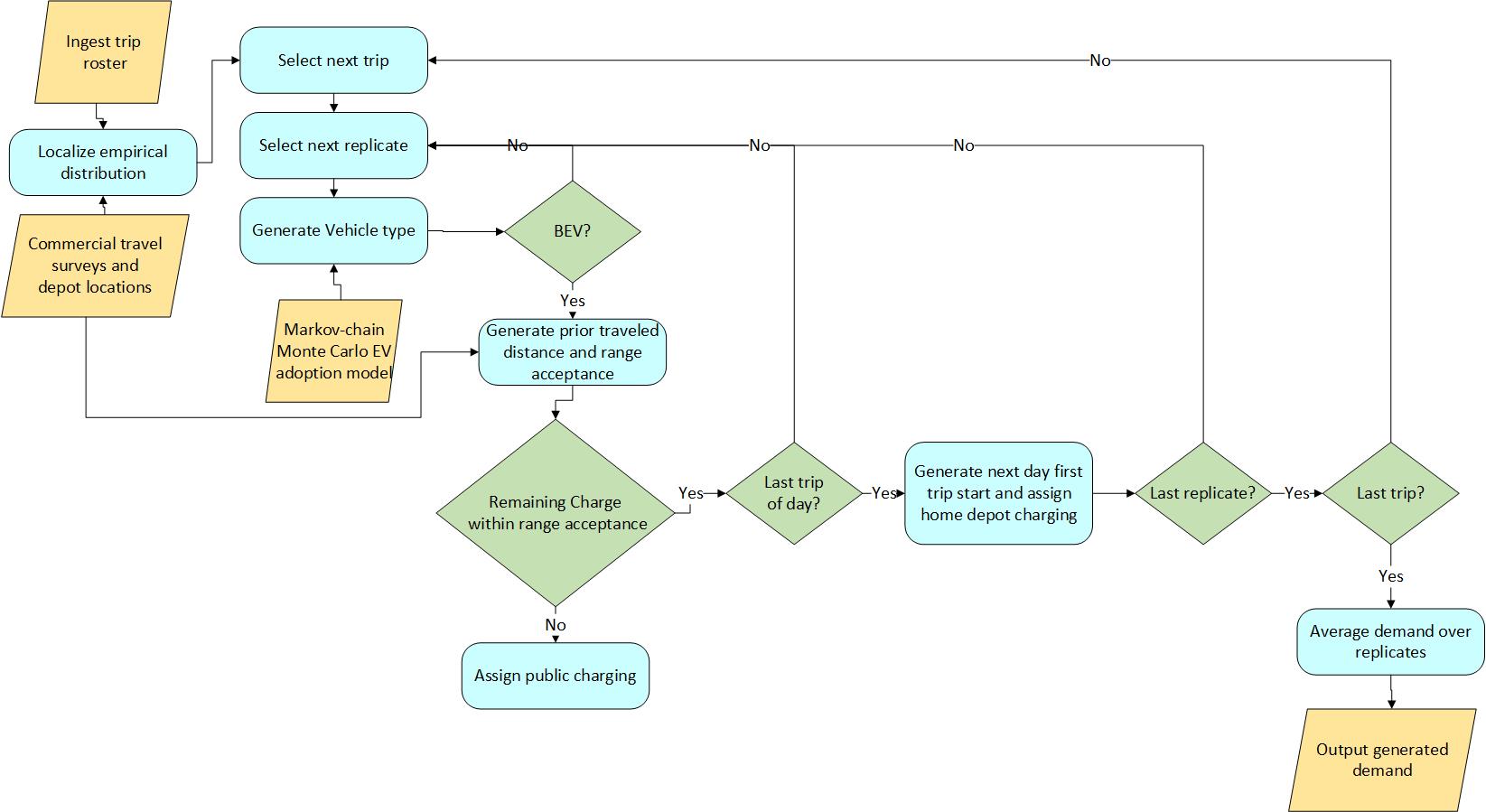}
    \caption{Framework for transportation and EV charging demand}
    \label{fig:transportframe}
\end{figure*}

The EV and traffic modeling for the methodology is described in further detail in \cite{annpaper}.

%% file: Sections/Mapping.tex
Substation or node service areas can be defined using Voronoi diagrams. Intuitively, a Voronoi diagram \cite{map64} can be understood as the diagram made with a given set of points, or "seeds," with expanding radii. When one seed’s expanded radius hits another, a boundary between those seeds’ areas is formed. Ultimately, the combined area of all seeds has expanded to encompass all eligible area, leaving a set of tessellating polygons remaining, each corresponding
to its initial seed.
Formally, a Voronoi diagram can be described as the union of the Voronoi regions \cite{map65}. A
Voronoi region is defined in Equation \ref{eq:211} for a point s in the set of total points (S). The complete
Voronoi diagram can be expressed as in Equation \ref{eq:212}. Voronoi diagrams are employed, for the mapping of EV chargers, due to the nature of the system. Since the system is synthetic, Voronoi diagrams are used to create synthetic distribution service areas served by each transmission substation.

\begin{equation} \label{eq:211}
    Vor(s)={p:distance(s,p)\leq distance(s',p), \forall s' \varepsilon S}
\end{equation}

\begin{equation} \label{eq:212}
    Vor(S)=\cup_{s \varepsilon S} Vor(s)
\end{equation}

This mapping can be performed at three levels of granularity, depending on studies to be
performed and information available including:
\begin{itemize}
    \item transmission-level mapping and simulation
    \item distribution-level mapping and simulation
    \item distribution-level mapping for transmission-level simulation.
\end{itemize}
The three levels of granularity rely on the same process with difference in the “seeds" used to create the Voronoi diagram.
The procedure is as follows:
\begin{itemize}
    \item Depending on the desired granularity of mapping, use the location of either transmission level substations or distribution-level nodes as starting points
    \item Create a Voronoi diagram,
    \item  If performing a distribution-level mapping for a transmission-level simulation:
    \begin{itemize}
        \item Identify Voronoi polygons corresponding to distribution feeders served by the selected
        substation
        \item Aggregate the Voronoi polygons to create a transmission-level substation service territory
        \item Repeat for all transmission-level substations.
        \item Using EV charger locations, map EV chargers to their corresponding electric grid service territory
        \item Add loads to power system models to represent EV chargers.
    \end{itemize}
\end{itemize}

The result of this mapping is tessellating service territories corresponding to transmission level substations or distribution-level nodes. The mapping only needs to be performed once for
a given region. If a substation has an EV charging node fall within its geographic footprint, it indicates that the charger's most proximate distribution point of interconnection would aggregate to the specified transmission-level substation and thus, its load is best represented as an addition to the identified transmission-level substation.

%% file: Sections/III_CoSimulation.tex
\subsection{Co-Simulation Framework}
The distribution and transmission networks, used in this study, are coordinated and simulated together, to discover optimal placement and sizing, for constructing EV charging stations. The ac OPF of the combined system is calculated by first calculating the ac OPF for the distribution network and then for the transmission system. Variables are shared between the two systems using the HELICS framework \cite{helics}. The voltage magnitude and angle, of the 69kV buses, alongside their real and reactive loads are the shared variable between transmission and distribution systems. These values from the results of the 3 phase unbalanced distribution system power flow are then used to initialize the ac OPF for the transmission network. At each time step the shared variables for both systems are sent to a controller which determines if the two systems have converged to a shared optimal solution. If the marginal costs at proposed charging station reaches one of the thresholds described in \ref{tab:mctable} the load is delayed or shed depending on the time-step, which is then solved again to determine the costs of the station with the reduced load. 

\begin{figure*}[htbp]
    \centering
    \includegraphics[width=2\columnwidth]{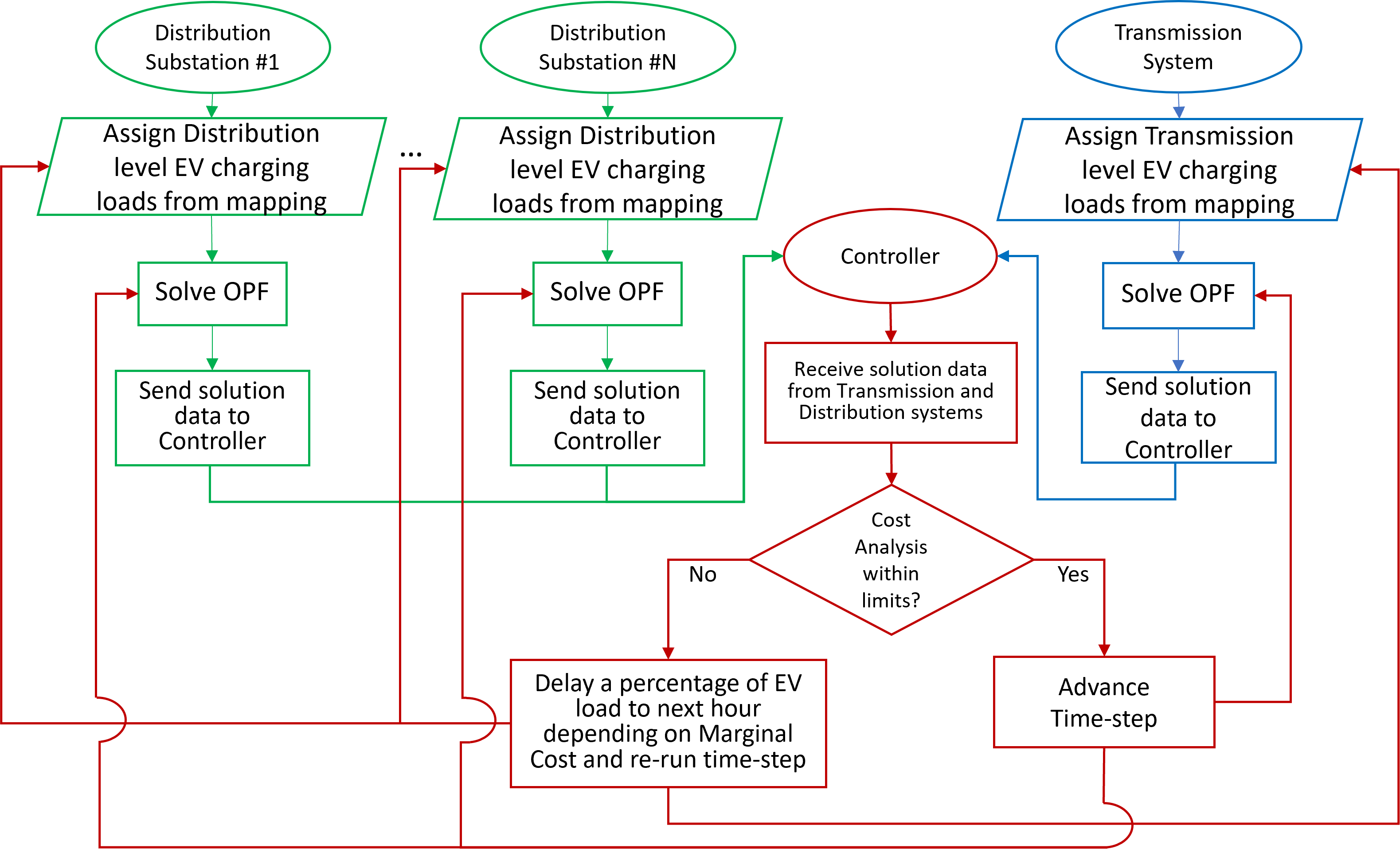}
    \caption{Co-Simulation framework for EV charging analysis}
    \label{fig:cosimframe}
\end{figure*}

The AC-OPF is solved in PowerWorld Simulator V23. PowerWorld Simulator solves the OPF by successive linear approximations, first solving the AC power flow then linearizing around that operating point, solving a linear programming (LP) formulation of the AC-OPF, then solving the AC power flow to ensure AC feasibility. This process repeats until either the LP AC-OPF and ACPF solutions differ less than a given threshold, or a maximum number of iterations is reached. Additionally, via bus voltage and line MVA limits are modeled as soft constraints with a penalty function in the objective function, allowing the PW Simulator AC-OPF solver to return an AC solution to an infeasible case, instead of simply returning an “infeasible” error like most commercial optimization solvers will do. This approach is much more useful to power system planners and operators instead of just seeing an “infeasible” error, with no indication about the performance of the rest of the system. 

\subsection{AC Optimal Power Flow (AC-OPF)}
AC-OPF \cite{wood2013power} is solved to determine the steady-state output power of generators, power flowing between the lines, and voltage outputs of buses in the distribution and transmission power system in a way to minimize the operation cost and satisfy the power grid constraints. Coefficients ($a$, $b$, and $c$) that represent quadratic cost curve elements of generators specifies $\mathcal{F}_c(P_G)$:

\begin{mini!}|s|[1]<break>
{P_G}
{\mathcal{F}_c(P_G) = \sum\limits_{g=1}^{|\mathcal{G}|} 
 [a_g + b_g P_G,g + c_g P^2_G,g]  \label{objective}}
{\label{eq:OF1}}
{}
\addConstraint{P_{G,(g\in g(i))} - P_{D,i} = |V_i|\sum\limits_{k=1}^{|N|}|V_k| (G^Y_{ik} cos \theta_{ik} + B^Y_{ik} sin \theta_{ik})\label
{eq:OF2}}
\addConstraint{Q_{G,(g\in g(i))} - Q_{D,i} = |V_i|\sum\limits_{k=1}^{|N|}|V_k| (G^Y_{ik} sin \theta_{ik} - B^Y_{ik} cos \theta_{ik}) \label{eq:OF3}}
\addConstraint{P_{min,g}  \le P_{G,g}  \le P_{max,g}        \qquad  \forall{g \in \mathcal{G}} \label{eq:OF4}}
\addConstraint{Q_{min,g}  \le Q_{G,g}  \le Q_{max,g}        \qquad  \forall{g \in \mathcal{G}}  \label{eq:OF5}}
\addConstraint{V_{min,i}  \le |V_i|  \le V_{max,i}        \qquad  \forall{i \in \mathcal{N}}   \label{eq:OF6}}
\addConstraint{P_l^2 + Q_l^2 \le S^2_{max,l} \qquad \forall{e\in \mathcal{E}}   \label{eq:OF7}}
\addConstraint{ P_l = |V_i|^2G^Y_{ik}-|V_i||V_k|(G^Y_{ik}cos \theta_{ik} + B^Y_{ik}sin \theta_{ik}) \label{eq:OF8}}
\addConstraint{Q_l = -|V_i|^2B^Y_{ik}-|V_i||V_k|(B^Y_{ik}cos \theta_{ik} - G^Y_{ik}sin \theta_{ik})  \label{eq:OF9}}
\end{mini!}

Equation (\ref{eq:OF1}) is the objective function of the ac OPF and the constraints including active and reactive power balance equations (\ref{eq:OF2}, \ref{eq:OF3})  as well as additional operational constraints equation from (\ref{eq:OF4}) to (\ref{eq:OF7}) should be satisfied. \cite{wood2013power}.

$|V_i|$, in the equations is the voltage magnitude at the $i$\textsuperscript{th} bus, and $\theta_i$ is the voltage angle at the $i$\textsuperscript{th} bus. The voltage angle difference betwen the $i$\textsuperscript{th} and $k$\textsuperscript{th} buses is the voltage angle diffence $\theta_{ik}$. $N$ is the number of buses in the system. $P_{D,i}$ and $Q_{D,i}$ are the real and reactive power demands at the $i$\textsuperscript{th} bus respectively. Similarly, $P_{G,g}$ and $Q_{G,g}$ are the real and reactive power generation of the $g$\textsuperscript{th} generator, respectively. It is noticeable that $\mathcal{G}$ is the amount of all generators in the system.
 The bus admittance matrix is expressed by as a real part $G^Y_{ik}$, and an imaginary part $B^Y_{ik}$. Maximum as well as minimum operating limits in the generator are supplied by  $(P_{min,g}, P_{max,g})$ for real power, and $(Q_{min,g}, Q_{max,g})$ for reactive power. $(V_{min,i}, V_{max,i})$ are limited in voltage magnitude of each bus. The power flow of the branch, $l$, is its thermal limit, also $S_{max,e}$ is involved in real and reactive power flow in equation (\ref{eq:OF7}). The power flow of the branches including lines and transformers in the grid are calculated in equations (\ref{eq:OF8}-\ref{eq:OF9}). It should be noted that $\mathcal{E}$ is the number of all branches in the power system.
 
\subsection{Network Constraints}
As mentioned in the previous section, Equations (\ref{eq:OF8}-\ref{eq:OF9}) are used to calculate the real and reactive power that flows between each connected pair of buses. Each of these branches has a limited apparent power transfer capacity described by Equation (\ref{eq:OF7}). 

In additon to the Equation (\ref{eq:OF7}) limit, we also take into account the North American Electric Reliability Cooperation (NERC) standards on system operating limit and
exceedance clarification. This standard applies to a 24 hour continuous line rating as a normal rating. The acceptable variations in acceptable over-limiting based on period of time overloaded and season are shown in \ref{tab:acceptablerating}. These ratings are based on the American National Standards Institute (ANSI) C37.010 limits.
Emergency ratings, generally 10\% of line capacity, typically last less than 24-hours, in winter they tend to last 4-hours, while in summer they can last  for 12-hours. Short-term emergency ratings can be up to 15\% of line capacity, but are usually only able to withstand a period of 15 minutes \cite{NREC_SOL},\cite{ISO-NE_CapacityRating}.

\begin{table}[]
\centering
\caption{\\Acceptable Multipliers of Nameplate Rating}
\label{tab:acceptablerating}
\begin{tabular}{ccc}
\hline
Ratings & Winter & Summer \\ \hline
Normal & 1.23 & 1.1 \\ 
Emergency – 15 minutes & 1.83 & 1.67 \\ 
Emergency – 4 hours & 1.34 & - \\ 
Emergency – 12 hours & - & 1.18 \\ \hline
\end{tabular}
\end{table}

The line overloads in the grid that are greater than the acceptable normal or emergency MVA ratings and last for a longer duration than the defined standards, can create major reliability issues.

\subsection{Marginal Cost Constraints}
The study was performed with two different methods of EV charging, one with and one without EV load flexibility, future work will more thoroughly detail the results from this comparison. With the method including load felxibility, to determine optimal charging patterns from a cost benefit perspective, marginal cost thresholds were established. When the marginal cost thresholds shown in Table \ref{tab:mctable} are reached a percentage of EV charging demand is delayed by applying a multiplier, Eq. \ref{eq:MC1}, and the excess load is shifted to the next hour, Eq. \ref{eq:MC2}. For the purposes of this study, the final hour, 11:00 PM, sheds load without assigning it to the next hour, if a threshold is met.

\begin{equation} \label{eq:MC1}
    EV_{hour} = EV_{hour}*M_{MC}
\end{equation}
if hour $<$ 24
\begin{equation} \label{eq:MC2}
    EV_{hour+1} = EV_{hour+1} + EV_{hour}*(1-M_{MC})
\end{equation}
Where $M_{MC}$ is the multiplier given to the load based on its Marginal Cost as described in Table \ref{tab:mctable}.

\begin{table}[]
\centering
\caption{Marginal Cost Thresholds for EV Charging Delay}
\label{tab:mctable}
\begin{tabular}{cccc}
\hline
\multirow{4}{*}{\begin{tabular}[c]{@{}l@{}}12:00 AM-\\ 10:00 PM\end{tabular}} &
  Marginal Cost &
  \begin{tabular}[c]{@{}l@{}}Percentage of\\ EV charging delayed\\ to next hour\end{tabular} &
  $M_{MC}$ \\ \cline{2-4} 
 & \textgreater{}\$500 & 80\% & .2 \\ \cline{2-4} 
 & \textgreater{}\$100 & 70\% & .3 \\ \cline{2-4} 
 & \textgreater{}\$60  & 50\% & .5 \\ \hline
\multirow{4}{*}{11:00 PM} &
  Marginal Cost &
  \begin{tabular}[c]{@{}l@{}}Percentage of\\ EV load shed\end{tabular} &
  $M_{MC}$ \\ \cline{2-4} 
 & \textgreater{}\$1000 & 90\% & .1 \\ \cline{2-4} 
 & \textgreater{}\$500 & 60\% & .4 \\ \cline{2-4} 
 & \textgreater{}\$100 & 30\% & .7 \\ \hline
\end{tabular}
\end{table}

%% file: Sections/IV_Case_Study.tex
EV integration studies on a synthetic grid are useful for several reasons. Although we cannot use a synthetic grid to examine impacts of changes on the real grid, we can test the performance and scalability of simulation, control and optimization algorithms on synthetic grids. Additionally,  comparing changes in operating cost, capital cost, or emissions to a base case can be done whether or not a case is synthetic and still have real world implications, since synthetic Grids have been shown to perform very similarly to real electric grids \cite{synthgrid}, \cite{synthgrid2}, \cite{synthgrid3}.
The combined transmission and distribution system used for this co-simulation was built and validated using the process outlined in \cite{newsynthpaper} and \cite{synthpaperorig}.

The 96 case studies were performed using large-scale modeling and simulation frameworks. These frameworks include a synthetic transmission model, based on the ERCOT footprint with 7,000 buses that encompasses Texas, and a distribution system consisting of three metropolitan areas, Houston and the combined Dallas Fort Worth area, and the area along the highway that connects the three cities, the I-45 corridor. The distribution system contains 6,566 distribution feeders with over 3,000,000 distribution system nodes connected to 1,841 transmission substations. 
This study is many orders of magnitude larger than most transmission and/or distribution research projects, which typically use less than 10 distribution feeders with less than 100 nodes per feeder

\begin{figure}[htbp]
    \centering
    \includegraphics[width=\columnwidth]{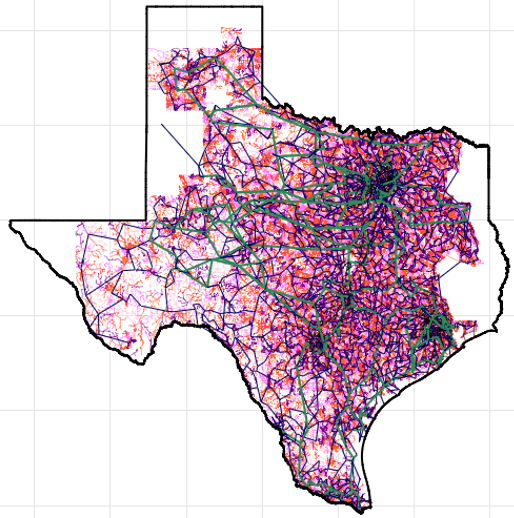}
    \caption{Texas system used in the case study}
    \label{fig:texas-system}
\end{figure}

\subsection{Distribution System}
The distribution system used for this study covers the areas encompassing two metropolitan areas, Dallas/Fort Worth and Houston along with the area around the highway that connects them, the I-45 corridor. The parameters of this system are shown in Table \ref{tab:dist-params}. An unbalanced 3-phase distribution power flow is performed to find distribution line and transformer overloads. 

\begin{table}[]
\centering
\caption{Distribution System Parameters}
\label{tab:dist-params}
\begin{tabular}{cc}
\hline
Attribute              & Value      \\ \hline
Number of Substations  & 1841       \\ \hline
Number of Feeders      & 6,566      \\ \hline
Cumulative feeder length (mi)       & 185,330           \\ \hline
Number of Line Segments        & 10,052,796 \\ \hline
Number of Transformers & 1,729,184  \\ \hline
Number of Nodes & 3,996,410  \\ \hline
\end{tabular}
\end{table}

\subsection{Transmission System}
The transmission system used in this study covers the entire Texas system. Using this model, we were able to minimizing transmission operational cost using positive sequence modeling, grid operational cost, power plant emissions, and Transmission overloads.

\begin{table}[]
\centering
\caption{Transmission System Parameters}
\label{tab:trans-param}

\begin{tabular}{cc}
\hline
Attribute          & Value \\ \hline
Buses              & 6,717 \\ \hline
Substations        & 4,894 \\ \hline
Areas              & 8     \\ \hline
Transmission lines & 7,173 \\ \hline
Transformers       & 1,967 \\ \hline
Loads              & 5,095 \\ \hline
Generators         & 731   \\ \hline
Shunts             & 634   \\ \hline
Peak load (GW)     & 75    \\ \hline
\end{tabular}%
\end{table}

\subsection{Transportation Data}
\begin{figure}[htbp]
    \centering
    \includegraphics[width=\columnwidth]{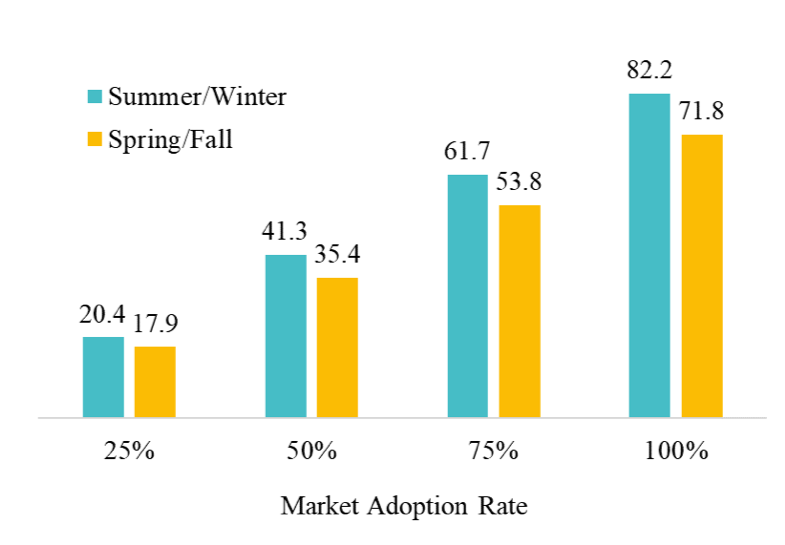}
    \caption{24-Hour Megaregion charging demand (GWh) by Market Adoption Rate; 200kW Charge Rate and I-45 Depot at Midpoint
}
    \label{fig:exampleoutput}
\end{figure}

\begin{figure*}[htbp]
    \centering
    \includegraphics[width=2.15\columnwidth]{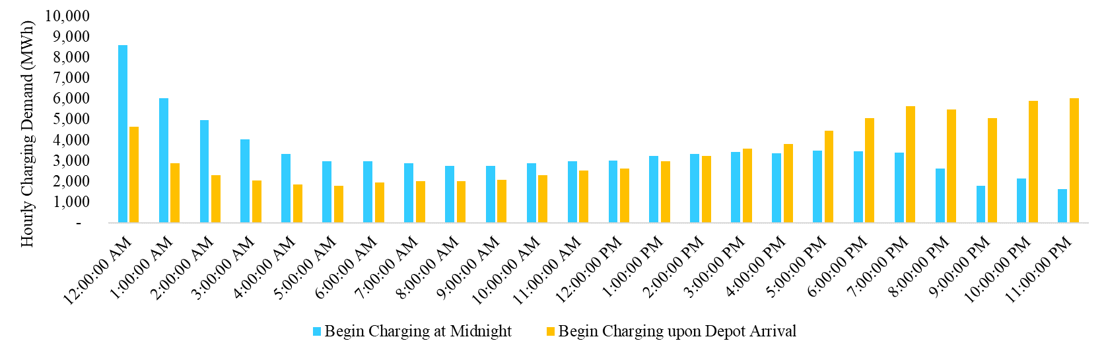}
    \caption{200kW Charge Rate, Winter/Summer AC Use, 100\% Market Adoption and Depot at I-287}
    \label{fig:chargerate}
\end{figure*}

The parameters for the 96 scenarios are shown in Table \ref{tab:params}. The scenarios vary the season, EV market adoption rate, charging logic, and charging locations to varying degrees depending on the scenario. Figure \ref{fig:exampleoutput} shows the market adoption rates during peak and shoulder seasons. Additionally, Fig. \ref{fig:chargerate} shows the difference in charging demand depending on when charging begins.

\begin{table}[]
\centering
\caption{Scenario Parameters}
\label{tab:params}
\begin{tabular}{cc}
\hline
Charge rate:             & 100, 200, 300 kW                   \\ \hline
Season:                  & Peak, Shoulder                     \\ \hline
EV market adoption rate: & 25\%, 50\%, 75\%, and 100\%        \\ \hline
Charging logic:          & upon arrival and start at midnight \\ \hline
Connecting Highway (I-45) charging location: & \begin{tabular}[c]{@{}c@{}}midpoint between \\Houston and Dallas\\  and close to Dallas\end{tabular} \\ \hline
\end{tabular}
\end{table}

%% file: Sections/V_Results_Discussion.tex
The results of the study show a decrease in emission with the adoption of EVs. The largest impact on grid-based emissions comes from the market adoption rate,  with the rate of emissions increase going down as the market adoption rate increases. This is most likely due to the fact that coal is cheaper to operate but more polluting than other methods. However, as EV load increases, natural gas power plants will be used more. Charging at midnight increases emissions, which increases concurrent load resulting in more polluting plants being dispatched.

\begin{figure}[htbp]
    \centering
    \includegraphics[width=\columnwidth]{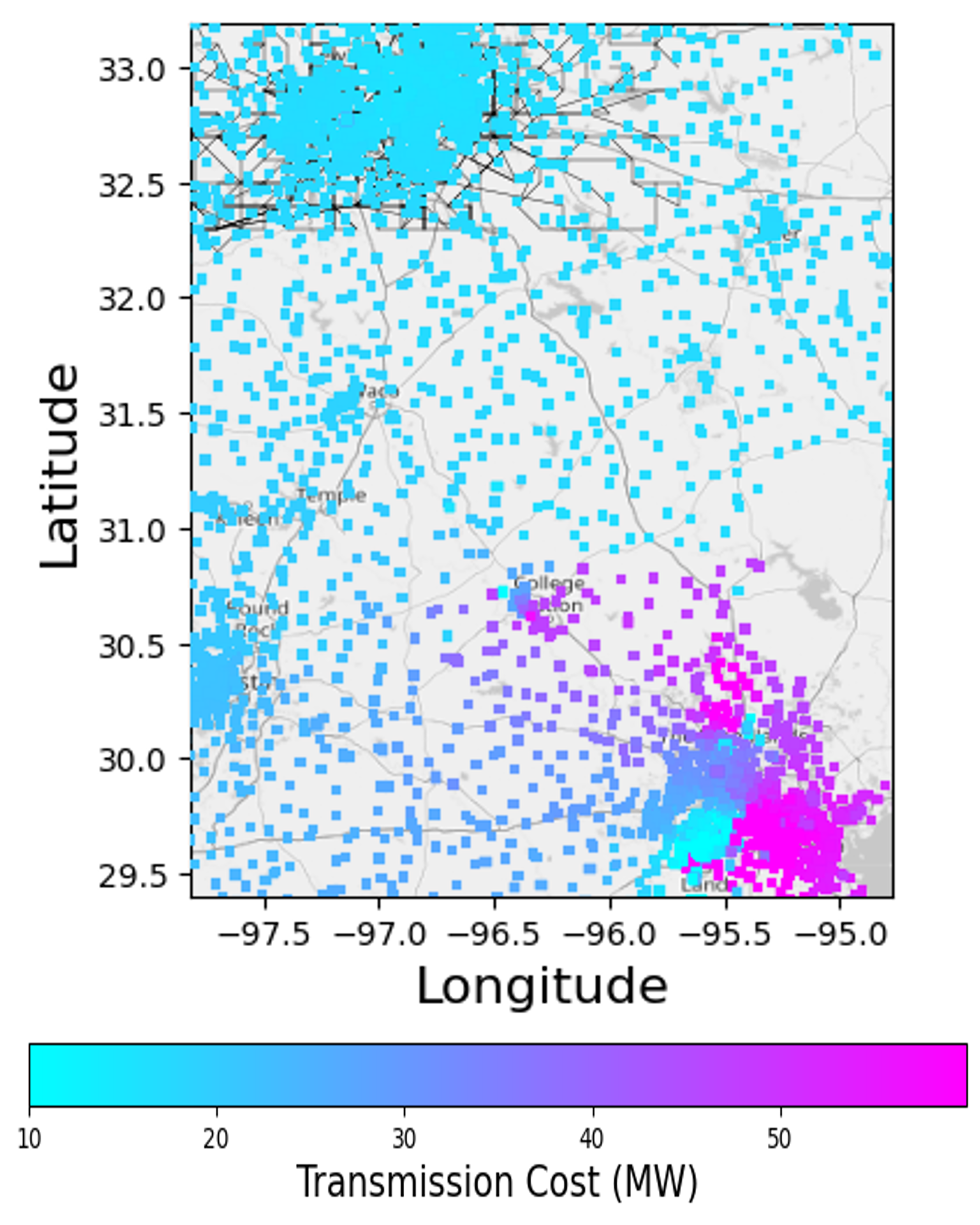}
    \caption{Transmission bus marginal costs at 12:00 AM}
    \label{fig:marginalcost}
\end{figure}

 As for cost savings and capital costs associated with EV charging, all 96 scenarios showed cost savings, which mirrors the savings from the operational cost. Delaying EV charging always resulted in cost savings, ranging from 79-100\% . This cost saving is the savings on transmission capital cost to remediate overloaded transmission lines and transformers. The LMP, of the buses, in the system is driven by overloaded and congested transmission lines. This makes it clear that delaying EV load at high LMP buses eliminates most overloads. A map showing the LMPs of individual buses at 12:00 AM is shown in Fig. \ref{fig:marginalcost}. This drop in LMP allows for a decrease in both transmission capital and operating costs. A summary of the capital cost across all scenarios showing this trend is shown in Table \ref{tab:capital_cost}.

\begin{table*}[]
\centering
\caption{Capital Cost Summary}
\label{tab:capital_cost}
\begin{adjustbox}{width=\textwidth}
\begin{tabular}{cccccc}
\hline
Charging Logic &
  Baseline Scenario Parameter &
  Comparison Parameter &
  \begin{tabular}[c]{@{}c@{}}Average Operating Cost \\ Difference (yearly)\end{tabular} &
  \begin{tabular}[c]{@{}c@{}}Average Transmission \\ Capital Cost\end{tabular} &
  \begin{tabular}[c]{@{}c@{}}Average Distribution \\ Capital Cost\end{tabular} \\ \hline
 &
  0\% adoption rate &
  25\% adoption rate &
  \$221,555,000 &
  \$6,460,000 &
  \$11,600,000 \\ \cline{2-6} 
 &
  25\% adoption rate &
  50\% adoption rate &
  \$260,975,000 &
  \$10,650,000 &
  \$8,720,000 \\ \cline{2-6} 
 &
  50\% adoption rate &
  75\% adoption rate &
  \$282,875,000 &
  \$12,470,000 &
  \$8,250,000 \\ \cline{2-6} 
\multirow{-4}{*}{Both averaged} &
  75\% adoption rate &
  100\% adoption rate &
  \$297,475,000 &
  \$12,950,000 &
  \$8,810,000 \\ \hline
  &
  Charging upon depot arrival &
  \begin{tabular}[c]{@{}c@{}}Charging beginning \\ at midnight\end{tabular} &
  {\color[HTML]{FF0000} (\$34,675,000)} &
  \$500,000 &
  \$9,370,000 \\ \hline
 &
  100 kW max charging rate &
  200 kW &
  (\$9,125,000) &
  \$2,020,000 &
  \$4,040,000 \\ \cline{2-6} 
\multirow{-2}{*}{Beginning at midnight} &
  200 kW max charging rate &
  300 kW &
  {\color[HTML]{FF0000} (\$12,045,000)} &
  {\color[HTML]{FF0000} (\$650,000)} &
  \$3,290,000 \\ \hline
 &
  100 kW max charging rate &
  200 kW &
  {\color[HTML]{FF0000} (\$730,000)} &
  \$670,000 &
  \$1,790,000 \\ \cline{2-6} 
\multirow{-2}{*}{Upon depot arrival} &
  200 kW max charging rate &
  300 kW &
  \$110,000 &
  \$420,000 &
  \$1,370,000 \\ \hline
\end{tabular}
\end{adjustbox}
\end{table*}

\begin{figure}[htbp]
    \centering
\includegraphics[width=\columnwidth]{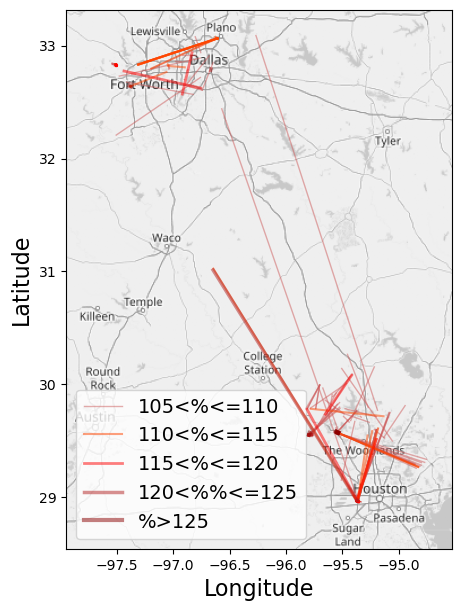}
    \caption{Locations and severity of overloaded lines at 12:00 AM}
    \label{fig:overloads}
\end{figure}

 The change in savings increases as market adoption rate increases. This occurs due to additional concurrent charging that overloads transmission lines. Charging at midnight achieves higher cost savings when compared with charging upon depot arrival over the course of a 24-hour simulation that starts at midnight. The rate structures had the same savings regardless of charging rate.
The overloaded lines found at 12:00 AM, during one of the 96 scenarios, scenario 13, is shown in Fig. \ref{fig:overloads}.

The rate structures provided in Table \ref{tab:mctable} result in small and inconsistent savings, or in some cases capital cost increases. This is due to our transmission and distribution co-simulation only including a OPF for the transmission system. This reflects the industry practice of having separate transmission and distribution operations centers that do not collaborate with each other in a meaningful way.

%% file: Sections/VII_Conclusion.tex
This paper proposed a unifying co-simulation infrastructure that integrates a variety of elements to study the effects of electric vehicle integration. Additionally the study provided results to interpret the optimal way to accelerate EV development and integration with cost optimization strategies to determine optimal placement and sizing of truck charging infrastructure that minimize infrastructure costs.
 
The strategy devised of strategically shifting load, to the next hour, when the LMP reaches a threshold both decreases costs and allows the system to operate with a lower overall load always reducing overloaded lines and often eliminating them. The 96 scenarios also showed that charging at midnight results in cost savings at all levels of integration with a reduction of up to 21\% in operation costs.

\section{Future Work}
Extending this work would involve creating a transmission/distribution co-optimization that focuses on operations, with optimal EV charging to mitigate the overloads foreseen in the distribution system. This would allow for even more precision in finding optimal placement of EV truck depots.
Future work should also modify the “charging beginning at midnight” scenario to start at 4 or 5pm and run 24 hours from then to better model how people charge their vehicles upon arriving home from work. Additionally, future work will use Power Models Integrated Transmission Distribution (PMITD) to do T\&D co-optimization.